\documentclass[aps,prd,preprint,superscriptaddress,showpacs]{revtex4}

\usepackage{graphicx}
\usepackage{dcolumn}
\usepackage{bm}

\def\simge{\mathrel{%
   \rlap{\raise 0.511ex \hbox{$>$}}{\lower 0.511ex \hbox{$\sim$}}}}
\def\simle{\mathrel{
   \rlap{\raise 0.511ex \hbox{$<$}}{\lower 0.511ex \hbox{$\sim$}}}}
\def\lessim{\mathrel {\vcenter {\baselineskip 0pt \kern 0pt
\hbox{$<$} \kern 0pt \hbox{$\sim$} }}}
\def\gessim{\mathrel {\vcenter {\baselineskip 0pt \kern 0pt
\hbox{$>$} \kern 0pt \hbox{$\sim$} }}}
\def \rightdownarrow
 {\kern.4em {\raise1.75ex\hbox{$\Bigl |$}$\kern-0.27em{\longrightarrow}$}}
\def \mrightdownarrow
 {\kern.4em {\raise1.25ex \hbox{$|$} \kern-0.27em{
                                                \longrightarrow}}}

\def\mMEt{\not\kern-.35em {E_T}}
\def\MEt{\hbox{$\mMEt$}}

\begin{document}




\title{Measurement of the Top Quark Mass in $p\bar{p}$
Collisions at $\sqrt{s} = 1.96$ TeV using the Decay Length
Technique}




\affiliation{Institute of Physics, Academia Sinica, Taipei, Taiwan 11529, Republic of China} 
\affiliation{Argonne National Laboratory, Argonne, Illinois 60439} 
\affiliation{Institut de Fisica d'Altes Energies, Universitat Autonoma de Barcelona, E-08193, Bellaterra (Barcelona), Spain} 
\affiliation{Baylor University, Waco, Texas  76798} 
\affiliation{Istituto Nazionale di Fisica Nucleare, University of Bologna, I-40127 Bologna, Italy} 
\affiliation{Brandeis University, Waltham, Massachusetts 02254} 
\affiliation{University of California, Davis, Davis, California  95616} 
\affiliation{University of California, Los Angeles, Los Angeles, California  90024} 
\affiliation{University of California, San Diego, La Jolla, California  92093} 
\affiliation{University of California, Santa Barbara, Santa Barbara, California 93106} 
\affiliation{Instituto de Fisica de Cantabria, CSIC-University of Cantabria, 39005 Santander, Spain} 
\affiliation{Carnegie Mellon University, Pittsburgh, PA  15213} 
\affiliation{Enrico Fermi Institute, University of Chicago, Chicago, Illinois 60637} 
\affiliation{Comenius University, 842 48 Bratislava, Slovakia; Institute of Experimental Physics, 040 01 Kosice, Slovakia} 
\affiliation{Joint Institute for Nuclear Research, RU-141980 Dubna, Russia} 
\affiliation{Duke University, Durham, North Carolina  27708} 
\affiliation{Fermi National Accelerator Laboratory, Batavia, Illinois 60510} 
\affiliation{University of Florida, Gainesville, Florida  32611} 
\affiliation{Laboratori Nazionali di Frascati, Istituto Nazionale di Fisica Nucleare, I-00044 Frascati, Italy} 
\affiliation{University of Geneva, CH-1211 Geneva 4, Switzerland} 
\affiliation{Glasgow University, Glasgow G12 8QQ, United Kingdom} 
\affiliation{Harvard University, Cambridge, Massachusetts 02138} 
\affiliation{Division of High Energy Physics, Department of Physics, University of Helsinki and Helsinki Institute of Physics, FIN-00014, Helsinki, Finland} 
\affiliation{University of Illinois, Urbana, Illinois 61801} 
\affiliation{The Johns Hopkins University, Baltimore, Maryland 21218} 
\affiliation{Institut f\"{u}r Experimentelle Kernphysik, Universit\"{a}t Karlsruhe, 76128 Karlsruhe, Germany} 
\affiliation{High Energy Accelerator Research Organization (KEK), Tsukuba, Ibaraki 305, Japan} 
\affiliation{Center for High Energy Physics: Kyungpook National University, Taegu 702-701, Korea; Seoul National University, Seoul 151-742, Korea; and SungKyunKwan University, Suwon 440-746, Korea} 
\affiliation{Ernest Orlando Lawrence Berkeley National Laboratory, Berkeley, California 94720} 
\affiliation{University of Liverpool, Liverpool L69 7ZE, United Kingdom} 
\affiliation{University College London, London WC1E 6BT, United Kingdom} 
\affiliation{Centro de Investigaciones Energeticas Medioambientales y Tecnologicas, E-28040 Madrid, Spain} 
\affiliation{Massachusetts Institute of Technology, Cambridge, Massachusetts  02139} 
\affiliation{Institute of Particle Physics: McGill University, Montr\'{e}al, Canada H3A~2T8; and University of Toronto, Toronto, Canada M5S~1A7} 
\affiliation{University of Michigan, Ann Arbor, Michigan 48109} 
\affiliation{Michigan State University, East Lansing, Michigan  48824} 
\affiliation{Institution for Theoretical and Experimental Physics, ITEP, Moscow 117259, Russia} 
\affiliation{University of New Mexico, Albuquerque, New Mexico 87131} 
\affiliation{Northwestern University, Evanston, Illinois  60208} 
\affiliation{The Ohio State University, Columbus, Ohio  43210} 
\affiliation{Okayama University, Okayama 700-8530, Japan} 
\affiliation{Osaka City University, Osaka 588, Japan} 
\affiliation{University of Oxford, Oxford OX1 3RH, United Kingdom} 
\affiliation{University of Padova, Istituto Nazionale di Fisica Nucleare, Sezione di Padova-Trento, I-35131 Padova, Italy} 
\affiliation{LPNHE, Universite Pierre et Marie Curie/IN2P3-CNRS, UMR7585, Paris, F-75252 France} 
\affiliation{University of Pennsylvania, Philadelphia, Pennsylvania 19104} 
\affiliation{Istituto Nazionale di Fisica Nucleare Pisa, Universities of Pisa, Siena and Scuola Normale Superiore, I-56127 Pisa, Italy} 
\affiliation{University of Pittsburgh, Pittsburgh, Pennsylvania 15260} 
\affiliation{Purdue University, West Lafayette, Indiana 47907} 
\affiliation{University of Rochester, Rochester, New York 14627} 
\affiliation{The Rockefeller University, New York, New York 10021} 
\affiliation{Istituto Nazionale di Fisica Nucleare, Sezione di Roma 1, University of Rome ``La Sapienza," I-00185 Roma, Italy} 
\affiliation{Rutgers University, Piscataway, New Jersey 08855} 
\affiliation{Texas A\&M University, College Station, Texas 77843} 
\affiliation{Istituto Nazionale di Fisica Nucleare, University of Trieste/\ Udine, Italy} 
\affiliation{University of Tsukuba, Tsukuba, Ibaraki 305, Japan} 
\affiliation{Tufts University, Medford, Massachusetts 02155} 
\affiliation{Waseda University, Tokyo 169, Japan} 
\affiliation{Wayne State University, Detroit, Michigan  48201} 
\affiliation{University of Wisconsin, Madison, Wisconsin 53706} 
\affiliation{Yale University, New Haven, Connecticut 06520} 
\author{A.~Abulencia}
\affiliation{University of Illinois, Urbana, Illinois 61801}
\author{J.~Adelman}
\affiliation{Enrico Fermi Institute, University of Chicago, Chicago, Illinois 60637}
\author{T.~Affolder}
\affiliation{University of California, Santa Barbara, Santa Barbara, California 93106}
\author{T.~Akimoto}
\affiliation{University of Tsukuba, Tsukuba, Ibaraki 305, Japan}
\author{M.G.~Albrow}
\affiliation{Fermi National Accelerator Laboratory, Batavia, Illinois 60510}
\author{D.~Ambrose}
\affiliation{Fermi National Accelerator Laboratory, Batavia, Illinois 60510}
\author{S.~Amerio}
\affiliation{University of Padova, Istituto Nazionale di Fisica Nucleare, Sezione di Padova-Trento, I-35131 Padova, Italy}
\author{D.~Amidei}
\affiliation{University of Michigan, Ann Arbor, Michigan 48109}
\author{A.~Anastassov}
\affiliation{Rutgers University, Piscataway, New Jersey 08855}
\author{K.~Anikeev}
\affiliation{Fermi National Accelerator Laboratory, Batavia, Illinois 60510}
% ...
\collaboration{CDF Collaboration\footnote{With visitors from $^a$University of Athens, 
$^b$University of Bristol, 
$^c$University Libre de Bruxelles, 
$^d$Cornell University, 
$^e$University of Cyprus, 
$^f$University of Dublin, 
$^g$University of Edinburgh, 
$^h$University of Heidelberg, 
$^i$Universidad Iberoamericana, 
$^j$University of Manchester, 
$^k$Nagasaki Institute of Applied Science, 
$^l$University de Oviedo, 
$^m$University of London, Queen Mary and Westfield College, 
$^n$Texas Tech University, 
$^o$IFIC(CSIC-Universitat de Valencia), 
}}
\noaffiliation

\date{\today}

\begin{abstract}
We report the first measurement of the top quark mass using the decay length technique in $p\bar{p}$ collisions at a center-of-mass energy of 1.96 TeV.  This technique uses the measured flight distance of the $b$ hadron to infer the mass of the top quark in lepton plus jets events with missing transverse energy.  It relies solely on tracking and avoids the jet energy scale uncertainty that is common to all other methods used so far.   We apply our novel method to a 695~pb$^{-1}$ data sample recorded by the CDF~II detector at Fermilab and extract a measurement of $m_t = 180.7^{~+15.5}_{~-13.4}~(stat.)~\pm~8.6~(syst.)$ GeV/$c^2$.  While the uncertainty of this result is larger than that of other measurements, the dominant uncertainties in the decay length technique are uncorrelated with those in other methods.  This result can help reduce the overall uncertainty when combined with other existing measurements of the top quark mass. 
\end{abstract}


\maketitle


\newpage
\section{\label{sec:Intro}Introduction}
A precise determination of the top quark mass ($m_t$) is an important goal of high-energy physics. 
The uncertainty on $m_t$ is a dominant uncertainty in 
global standard model (SM) fits for the mass of the unobserved Higgs boson.  A precision measurement of $m_t$ constrains the allowed Higgs mass values within the SM.  It will tell us where to look for the Higgs and test whether it is the predicted SM particle or not after a signal has been established.  Recently, significant progress has been made in reducing 
the uncertainty in measurements of $m_t$~\cite{Abulencia:2005ak}.  Unfortunately, 
the most precise of the currently employed techniques are all limited by the same 
systematic uncertainty, the calorimeter jet energy scale.  

Some of the authors of this paper have developed a novel method to measure $m_t$ 
using the transverse decay length of $b$ hadrons from top decays~\cite{Hill:2005zy}.  
The method exploits the fact that top quarks produced in $p\bar{p}$ collisions at $\sqrt{s} = 1.96$~TeV are produced nearly at rest and decay almost instantaneously~\cite{toplife} to a relatively light bottom quark and a much heavier $W$ boson.  In the rest frame of the top quark, the relativistic boost given to the bottom quark as a
consequence of the top quark decay can be written simply as follows:

\begin{equation}
\gamma_b = \frac{m_{t}^{2} + m_{b}^{2} - m_{W}^{2}}{2 m_t m_b} \approx 0.4 \frac {m_t}{m_b}
\end{equation}

\noindent where $\gamma_b \equiv [1-(v_b^2/c^2)]^{-1/2}$ and the approximation makes use of the fact that $m_t \gg m_b$.  The mass of the top quark, therefore, is 
strongly correlated with the velocity imparted to the $b$ quark and the subsequent $b$ hadron after 
fragmentation.  Thus, the average momenta of the $b$ hadrons from top decays can 
be used to infer the mass of the top quark.  In this analysis, rather than measuring 
the average momenta, we simply measure the highly-correlated average 
transverse decay length of the $b$ hadrons, which we denote $\langle L_{2D} \rangle$. Furthermore,
we do not analytically solve for $m_t$ from $\langle L_{2D} \rangle$, but as detailed in Section~\ref{method} we establish the functional dependence of $m_t$ on $\langle L_{2D} \rangle$ using Monte Carlo (MC) simulations of signal and background events.

This technique relies on track reconstruction to determine precisely the decay length.  The calorimeter information is used only for the selection of event candidates.  Consequently, the uncertainty on the measurement due to the jet energy scale is negligible.  In this paper, we present the first measurement of $m_t$ using the decay length technique.  We apply this new method to $p\bar{p}$ collision data which were recorded by the CDF~II detector during Run~II of the Fermilab Tevatron.

\section{\label{sec:Detector} Detector Description }
The CDF~II detector is described 
in detail elsewhere~\cite{Acosta:2004yw}. The detector has a charged particle tracking
system immersed in a 1.4~T solenoidal magnetic field coaxial with the
$p$ and $\bar p$\ beams. Tracking over the radial range 1.5~cm to 28~cm is provided by three complementary silicon micro-strip detectors~\cite{Hill:2003ax,Sill:2000zz,Affolder:2000tj}.  A 3.1-m-long open-cell drift chamber covers the radial range from 40~to~137 cm~\cite{Affolder:2003ep}.  The
fiducial region of the silicon system extends to pseudorapidity~\cite{coord} of $|\eta| \sim2$, while the drift chamber (COT) provides coverage for $|\eta| \lesssim 1$.  Segmented electromagnetic and
hadronic calorimeters~\cite{Balka:1987ty,Bertolucci:1987zn,Albrow:2001jw} surround the tracking system and measure the energy of interacting particles in the pseudorapidity range $|\eta|<3.6$.  A set of drift chambers and scintillation counters~\cite{Ascoli:1987av} located outside the hadron calorimeter and another set behind a 60 cm iron shield detect muon candidates with $|\eta|<0.6$.  Additional chambers and counters detect muon candidates in the region $0.6 < |\eta| < 1.0$.  Cherenkov counters~\cite{Acosta:2002hx} located in the $3.7 < |\eta| < 4.7$ region measure the average number of inelastic $p\bar{p}$ collisions per bunch crossing and thereby determine the beam luminosity.

\section{\label{sec:Data}Data Sample and Event Selection}
In Run II of the Tevatron, protons and antiprotons collide at a center-of-mass energy of 1.96 TeV.
In such collisions, the SM predicts that $\sim85\%$ of $t\bar{t}$ pairs are produced through quark antiquark annihilation and $\sim15$\% are produced by gluon fusion. Top quarks are expected to decay almost exclusively to a $W$ boson and a $b$ quark.  The $W$ subsequently decays either hadronically, to a pair of quarks or leptonically, to a lepton and a neutrino. The final state $b\bar{b}\ell\bar{\nu}q\bar{q}^{\prime}$ (where $\ell = e,\mu$ only) resulting from one of each type of $W$ decay is called the ``lepton + jets" mode.  This channel has a large branching fraction with a good signal-to-background ratio; we use it to measure $m_t$ using the decay length technique.  Lepton + jets $t\bar{t}$ events typically contain a high transverse momentum ($p_T$) electron or muon, missing transverse energy ($\MEt$) from the undetected neutrino, and four high transverse energy ($E_T$) jets, two of which originate from $b$ quarks.  All methods employed in previous measurements of $m_t$ require the presence of all four jets, since they attempt to fully reconstruct the decays of both top quarks.  Sometimes one or more of the jets may not be reconstructed, making those events ineligible for traditional methods.  The decay length technique, however, can be applied to such events, providing the only measurement of $m_t$ from these data.

Results reported here are obtained from an analysis of data collected 
between March 2002 and September 2005. The data sample corresponds to an integrated luminosity of $\sim695$~pb$^{-1}$.  CDF~II employs a three-tiered trigger system to sequentially reduce event rates from $\sim1.7$~MHz to $\sim80$~Hz. The data used in this analysis were recorded with an inclusive lepton trigger that requires an electron (muon) with $E_T>$18~GeV  ($p_T>$18~GeV/$c$).  

From this inclusive lepton dataset we select events offline
with an electron (muon) with $E_T>$ 20~GeV ($p_T>$  20~GeV/$c$),
$\MEt>$20~GeV, and at least 3 jets with $|\eta|<$ 2 and energy-corrected~\cite{jetcorr} 
$E_T>$15~GeV~\cite{Bhatti:2005ai}. Electron candidates are required to have a well-measured track 
pointing at an energy deposit in the calorimeter.  The energy signature must be isolated 
from other calorimeter activity, and must have shower profiles consistent with expectation. 
We select muon candidates by requiring that the hadronic and electromagnetic energy deposited by the candidate in the calorimeter be consistent with that of a minimum ionizing particle. In addition, we match partially reconstructed tracks in the muon
chambers with well-measured tracks reconstructed in the COT. Jets are clustered using a fixed-cone algorithm with a cone size $\Delta R\equiv\sqrt{(\Delta\eta)^2 + (\Delta\phi)^2}= 0.4$.  Finally, in order to better distinguish $t\bar{t}$ events from background processes, we require at least one jet in the event to be identified as a $b$ jet (``tagged") by the reconstruction of a secondary vertex within that jet as described in Section~\ref{sec:LXY}.  We refer to the dataset selected above as the ``tagged lepton + jets" sample.

\section{\label{sec:LXY}Measurement of Transverse Decay Length}

The primary ({\sc primevtx}) and secondary vertex algorithms ({\sc secvtx}) used are described in Ref.~\cite{Abulencia:2006in}. {\sc primevtx} reconstructs the primary event vertex with a precision of $\sim15$~$\mu$m in the plane transverse to the beam for $t\bar{t}$ events.  {\sc secvtx} exploits the relatively long lifetime of $b$ hadrons in top decays to reconstruct a secondary vertex significantly displaced from the primary interaction. Secondary vertex tagging operates on a per-jet basis, where only tracks associated with the jet are considered.  We require that these tracks have at least three silicon hits attached to them, are within 2.0~cm from the primary vertex in the longitudinal direction, and that the final track fits have $\chi^2/d.o.f. \le 8.0$.  We select tracks contained inside a jet which are displaced with respect to the primary vertex if they have a large well-measured impact parameter with respect to that same vertex.  The {\sc secvtx} algorithm uses a two-pass approach to find secondary vertices from these selected tracks.  In the first pass, it attempts to reconstruct a secondary vertex which includes at least three tracks.  If the first pass is unsuccessful, a second pass is attempted which makes tighter track requirements and tries to reconstruct a two-track vertex.

Once a secondary vertex is found, we calculate the two-dimensional decay length, $L_{2D}$, as the projection onto the jet axis, in the plane transverse to the beam, of the vector pointing from the primary vertex to the secondary vertex.  The sign of $L_{2D}$ is given by the $\phi$ difference between the jet axis and the secondary vertex vector (positive if less than 90 degrees, negative if greater than 90 degrees).  The secondary vertices corresponding to the decay of $b$ and $c$ hadrons have large positive $L_{2D}$, while the secondary vertices from mismeasured tracks form a Gaussian distribution centered around $L_{2D} = 0$ with a width corresponding to the detector resolution.  A jet is ``positively tagged" if its transverse decay length divided by the uncertainty on that measurement ($L_{2D}/\sigma_{L_{2D}}$) is greater than 7.5.  Similarly, a jet is  ``negatively tagged" if $L_{2D}/\sigma_{L_{2D}} < -7.5$.  The positively tagged jet sample is predominantly composed of heavy-flavor ($b$ or $c$) jets while the negatively-tagged sample is mostly composed of light quark jets.

\begin{figure}[ht]
        \includegraphics[width=0.8\textwidth,clip=]{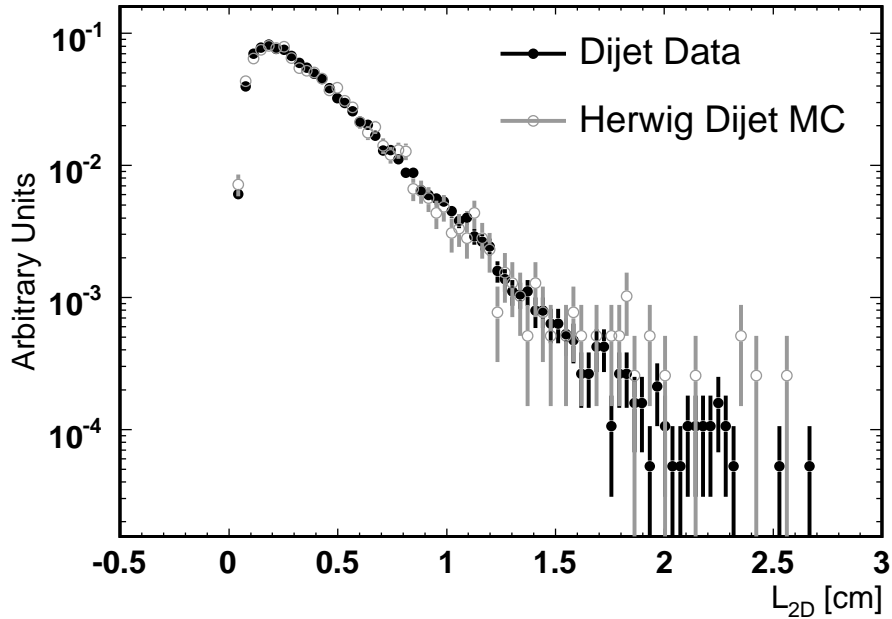}
    \caption{Comparison of $L_{2D}$ of positive tags in events with two {\sc secvtx} tags and an identified electron or muon in one of the two tagged jets.  Observed data and dijet MC are compared for this essentially triple tagged sample.  Both the lepton tag and the non-lepton tag are included.  Both distributions are normalized to unit area.\label{pos_tags}}
\end{figure}

This analysis requires an accurate simulation of $L_{2D}$.  To check the accuracy of the CDF~II simulation we examine heavy-flavor enriched data samples. We use dijet data samples recorded with a trigger requiring an 8 GeV electron (muon).  The lepton in these data, which we call a ``soft lepton," often comes from the semi-leptonic decay of a $b$ or $c$ quark such that the heavy-flavor content of these samples is enhanced relative to generic dijet  data.   We compare these to {\sc herwig}~\cite{Corcella:2000bw} generic dijet MC samples which have been pre-selected to contain a soft lepton in analogy with the trigger requirements imposed upon the observed data.  To increase the $b\bar{b}$ purity of these samples, we require that the soft lepton that triggered the event be contained within a jet with $E_T > 9$~GeV that is also positively tagged by {\sc secvtx}.  We also require the presence of another jet with $E_T > 9$~GeV, at least 2.0 radians away in azimuth, that is also positively tagged.   Finally, in order to distinguish $b$ quarks from $c$ quarks, we require the invariant mass of the four-vectors forming the secondary vertex~\cite{Abulencia:2006ce} of the tagged jets to be greater than 1.5 GeV/$c^2$.  With this essentially triple-tagged (two {\sc secvtx} plus one soft lepton) selection, a purity of $\sim99\%~b\bar{b}$ is obtained~\cite{purity}.  For all events passing the selection criteria, we make a histogram of the $L_{2D}$ for all positive tags.  These histograms are shown with observed data and MC overlaid in Fig.~\ref{pos_tags}.  We observe that the simulation models the $L_{2D}$ distribution very well.  We quantify this agreement by comparing $\langle L_{2D}\rangle$ for both observed data and MC as follows, where the errors are statistical only:

\begin{equation} 
	\langle L_{2D}^{data} \rangle = 0.378 \pm 0.002~\textrm{cm},~ \langle L_{2D}^{MC} \rangle = 0.381 \pm 0.004~\textrm{cm}
 \end{equation}   

\noindent From these, we compute a data/MC scale-factor which could be applied to the mean transverse decay length measured in the observed data of $0.992 \pm 0.012$.  This number is consistent with a scale factor of unity; we conclude that our simulation models the transverse decay length of $b$ hadrons with sufficient accuracy and do not apply any correction to the observed data.  This ratio encompasses many different possible sources of discrepancy between our observed data and our MC simulation including effects from detector resolution, fragmentation, and the relative proportions and lifetimes of the various $b$ hadrons.  As such, it is a comprehensive data-driven means of quantifying any systematic uncertainties in the measurement of $\langle L_{2D}\rangle$.  We note that in order to apply this treatment to $b$ hadrons in $t\bar{t}$ events, we rely on the assumption of the universality of $b$-fragmentation, {\it i.e.} that the fragmentation of a $b$ quark is independent of the process in which that $b$ quark was produced.  This assumption is predicted by the QCD factorization theorem~\cite{Ellis:1978sf} and is supported by a significant body of experimental evidence~\cite{Kniehl:2000hk}.  We assign the 1.2\% statistical uncertainty on the central value in the above calculation as a systematic uncertainty on the accuracy of our MC simulation. 

  We perform several additional checks on the data/MC scale-factor.  Since the average energy of $b$ hadrons from top decays is higher than that of those used to compute the scale factor, we examine the ratio as a function of jet $E_T$.  We find the scale factor to be independent of jet energy within uncertainties.  We also compute a data/MC ratio of the  $\langle L_{2D}\rangle$ for negatively tagged jets, thereby more directly checking the resolution modeling of the simulation.  We again measure a scale factor of unity within uncertainties.

\section{Sample Composition}
The tagged lepton + jets sample selected as described in Section~\ref{sec:LXY} has an 
expected signal-to-background ratio of about 2.5:1. The dominant background is the production of $W$ plus multijet events.  These events enter the signal sample when one of the jets is a $b$ jet or $c$ jet, or a light quark jet that has been mistakenly tagged as containing a secondary vertex.  We call the latter type of events ``mistags."  The other substantial background comes from collisions which do not produce a $W$ boson, termed  ``non-$W$" events.  These events are typically QCD multi-jet events where one jet has been misidentified as a high-$p_T$ lepton and mismeasured energies produce apparent $\MEt$.  Additionally, other processes such as $WW,WZ,ZZ,Z\rightarrow \tau\tau$, and single top contribute small amounts to the tagged lepton + jet sample.  The techniques used to calculate the expected contributions to this sample are detailed elsewhere~\cite{Abulencia:2006in}.  Estimated contributions to the tagged lepton + jets sample are summarized in Table~\ref{tab:bkg}. 

\begin{table}[t]
\begin{center}
\caption{\label{tab:bkg} Estimated number of events from background sources and single top that contribute to the tagged lepton + jets sample.  The number of events observed data is also presented.  The excess above the total background plus single top is assumed to be $t\bar{t}$ events.  Errors include statistical and systematic uncertainties.}
\begin{ruledtabular}
\begin{tabular}{cc} 
Source &  Number of Events\\
\hline
$Wb\bar{b}$~& $27.9 \pm 6. 2$\\
$Wc\bar{c}$~&  $12.2 \pm 3.2$\\
$Wc$ & $6.9 \pm 1.6$\\
non-$W$ & $12.5 \pm 2.6$\\
Mistags &$40.9 \pm 3.8$\\
$WW,WZ,ZZ,Z \rightarrow \tau \tau$ & $5.7 \pm 1.0$\\
Total Background & $106.1 \pm 10.5$\\
\hline
Single Top & $5.3 \pm 0.5$\\
Total Background + Single Top  & $111.4 \pm 11.0$\\
\hline
Data & 375\\
\end{tabular}
\end{ruledtabular}
\end{center}
\end{table}

\section{Expected $L_{2D}$ Distributions}
We generate {\sc herwig} $t\bar{t}$ MC samples using the {\sc cteq5l} parton distribution
functions~\cite{Lai:1999wy} followed by a detailed simulation of the CDF~II detector.  The CLEO {\sc qq} Monte Carlo simulation models the decays of $b$ and $c$ hadrons.  We produce these samples with top quark masses ranging from 130-230 GeV/$c^2$ in 5 GeV/$c^2$ intervals.   We subject these simulated events to the identical event selection as that required of the observed data.  After selection, we construct histograms of the transverse decay lengths of all positive tags in order to obtain $L_{2D}$ distributions for each mass point.  A similar process is performed for each of the backgrounds described above.  We model the $L_{2D}$ distributions for the $Wb\bar{b}$, $Wc\bar{c}$, and $Wc$ backgrounds using {\sc alpgen~\cite{Mangano:2002ea}} matrix element calculations which have been interfaced with {\sc herwig} to simulate the hadronization process.  To model the $L_{2D}$ distribution from the $W$ + mistag background we construct a hybrid data/MC template.  Mistags can either come from tracks which appear displaced due to limited impact-parameter resolution or from tracks that originate from actual long-lived particles that are not $b$ hadrons.  We use tags from negatively tagged $W$ + jets data, reflected about  $L_{2D} = 0$, to obtain a data-driven positive mistag shape to model the resolution contribution to the distribution.  We combine this with {\sc alpgen} interfaced to {\sc herwig} MC simulations of $W$ + multijets, which we rely on to model the contribution from long-lived particles such as $K_S$ and $\Lambda$.  We obtain the relative normalization for this combination from independent studies comparing positive and negative tags~\cite{Abulencia:2006in}.  For the purpose of this analysis, $WW,WZ,ZZ,Z \rightarrow \tau \tau$ events are considered mistags, and the mistag $L_{2D}$ distribution obtained above is used to model their small contribution.  We obtain the non-$W$ background $L_{2D}$ distribution directly from our observed data.  We select events with identical criteria to those for the signal sample, except the requirement that the lepton be calorimetrically isolated, where instead we explicitly require the lepton be non-isolated.  For most top analyses, single top is a background to the pair-produced signal.  With the decay length technique, however, this is not the case.  Although the correlation is not quite as strong as for pair-produced top quarks~\cite{singletop},  the $\langle L_{2D} \rangle$ from $b$ hadrons from single top decays is also correlated with $m_t$.  We use {\sc pythia}~\cite{Sjostrand:2000wi} MC to model the single-top $L_{2D}$ distribution as a function of top quark mass from 130-230 GeV/$c^2$.

\begin{figure}[ht]
  \begin{center}
    \includegraphics[width=0.8\textwidth,clip=]{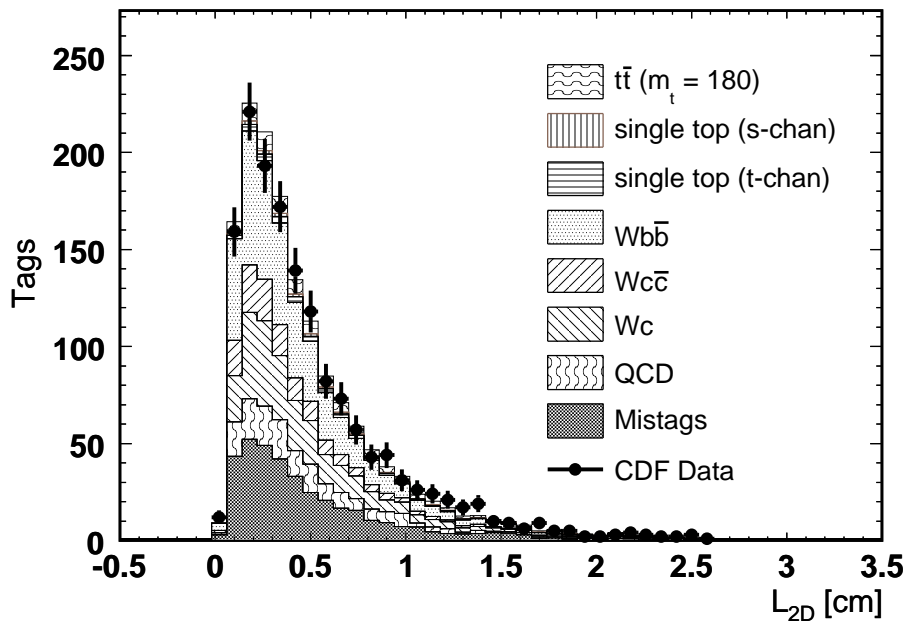}
    \caption{The $L_{2D}$ distribution of positive tags in lepton + jet events with 1 or 2 jets, for which only a small top contribution is expected.  The points are the observed data.  Expected contributions from signal and background MC are displayed cumulatively in the histogram.  To facilitate the shape comparison, the MC is normalized to the observed data.\label{data_bkgnd}}
  \end{center}
\end{figure}

As a cross-check on the modeling of $L_{2D}$ distributions for the various background processes, we examine the observed data in the background-dominated one- and two-jet events in the lepton + jet sample.
The $L_{2D}$ distribution of positive tags in selected events from the one and two jet bins is shown together with expected signal and background contributions in Fig.~\ref{data_bkgnd}.  Good agreement between MC and experimental data is observed; a Kolgomorov-Smirnov (KS) test yields a p-value of 30.6\%. 

\section{Estimation of Top Mass Dependence on $L_{2D}$ \label{method}}
We treat the signal and background $L_{2D}$ distributions described in the previous section as probability density functions from which we form ensembles of simulated experiments.  In forming each ensemble, the number of events from a given background source (and single top) is obtained by allowing the number of events for each process to fluctuate separately about the expected contributions listed in Table~\ref{tab:bkg}.  The number of events from $t\bar{t}$, which is similarly allowed to fluctuate, is taken to be the excess of the observed data in the tagged lepton + jets sample over the summed contributions of the background processes and single top production.  

In computing $\langle L_{2D}\rangle$, we will sum over tags rather than events.  We convert the number of events for each process to a number of tags by multiplying by the probability, obtained from MC simulation, for that process to contain more than one {\sc secvtx} tags.  We repeat this procedure 1,000 times for each mass point over the full mass range 130-230 GeV/$c^2$.  We construct histograms of the $\langle L_{2D}\rangle$ that results from each pseudo-experiment performed at a given $m_t$.  We extract the mean and $\pm1\sigma$ variance from these histograms, for each value of $m_t$, and fit these points to third-degree polynomials.  The fit to the mean establishes the most probable value for a true top mass given a measured $\langle L_{2D}\rangle$ and is the function used to make the top mass measurement from the $\langle L_{2D}\rangle$ observed in the data.   Similarly, the fits to the variance form $\pm1\sigma$ Neyman~\cite{Yao:2006px} confidence intervals which will be used to give the statistical uncertainty of the measurement in Section~\ref{results} as shown in Fig.~\ref{result1}.  

We derived these functions and validated our method prior to examining our experimental data.  We employed simulated data ensembles containing MC $t\bar{t}$ events with unknown top quark mass to demonstrate that $m_t$ could be extracted accurately and with appropriate precision. 

\section{Estimates of Systematic Uncertainties\label{sys}}

The systematic uncertainties for this measurement come from three kinds of sources.  The first arises from the accuracy of the modeling of factors which affect the top (or subsequent bottom) quark's momentum.  We estimate the uncertainty due to initial- and final-state gluon radiation by varying the relevant parameters by $\pm1$ standard deviation in the simulation~\cite{Abulencia:2005aj} and observing the effect on the measured $m_t$.  We quote half the difference between these variations as the systematic uncertainty.  To assess the systematic uncertainty from our choice of the {\sc cteq5l} parton distribution function, we observe the shift in $m_t$ that results when we substitute the {\sc mrst72} and {\sc mrst75} sets~\cite{Martin:2001es} which are evaluated at different values of the strong coupling constant, $\alpha_s$.  Additionally, we vary the 20 eigenvectors in the {\sc cteq6m} package~\cite{Pumplin:2002vw} by $\pm1$ standard deviation.  We add the shifts observed in $m_t$ in quadrature to determine the total systematic uncertainty.  We estimate the systematic uncertainty due to our choice of {\sc herwig} to simulate our signal events by measuring the shift induced in $m_t$ upon substitution of {\sc pythia} simulations.  Finally, we evaluate the uncertainty due to the jet energy scale by applying jet energy corrections which have been shifted by $\pm1$ standard deviations~\cite{Bhatti:2005ai} and taking half the difference as the systematic error.  As anticipated, we find the analysis to have negligible sensitivity to such shifts.

 The second type of systematic uncertainty comes from potential inaccuracies in the size or shape of background $L_{2D}$ distributions.  We quantify the uncertainty due to background normalization by  increasing/decreasing the contributions from each background process according to the uncertainties listed in Table~\ref{tab:bkg}.  We estimate the effects of uncertainties in the shape of the background $L_{2D}$ distributions by substituting altered distributions and noting the corresponding shift in $m_t$.  For the $W$ + heavy-flavor background shapes we vary the momentum transfer parameter ($q^2$) in the calculation.  For the non-$W$ shape we substitute a distribution obtained from observed $b\bar{b}$ data.  Finally, to estimate the systematic uncertainty due to the distribution used to model the mistag background, we measure the effect on $m_t$  when we alternatively use, only the shape derived from the observed data, and only the shape derived from MC, in place of the data/MC hybrid distribution used in the analysis. We take half the difference between these two determinations as the systematic uncertainty.  We add the separate background shape uncertainties in quadrature to arrive at a total systematic uncertainty.
 
 The final type of systematic uncertainty comes from imperfections of detector simulation of $L_{2D}$ or other, experimentally indistinguishable, disagreements between the $\langle L_{2D}\rangle$ in MC and observed data that may arise from inaccuracies in the simulation of $b$ hadron decays.  This uncertainty is taken as the error on the $\langle L_{2D}\rangle$ data/MC ratio as discussed in Section~\ref{sec:LXY}.  Table~\ref{tab:sys} lists each of the sources of systematic error and their corresponding uncertainties.  A total systematic uncertainty of 8.6 GeV/$c^2$ is assigned to the measurement.  The dominant sources of systematic error are the finite statistics used to derive the $\langle L_{2D}\rangle$ data/MC ratio and the modeling of the $L_{2D}$ distribution from mistags.

\begin{table}[t]
\caption{\label{tab:sys}
Summary of sources of systematic error and their estimated uncertainties.}
\begin{ruledtabular}
\begin{tabular}{cc} 
Source of Systematic Error & Uncertainty (GeV/$c^2$) \\
\hline
Monte Carlo Generator & 0.7 \\
Initial State Gluon Radiation & 1.0 \\
Final State Gluon Radiation & 0.9 \\
Parton Distribution Functions & 0.5 \\
Event Selection (Jet Energy Scale) & 0.3\\
Background Shape & 6.8 \\
Background Normalization & 2.3 \\
Multiple Interactions & 0.2 \\
Data/MC $\langle L_{2D} \rangle $~Ratio  & 4.2 \\
\hline
Total  & 8.6 \\
\end{tabular}
\end{ruledtabular}
\end{table}

\section{Results\label{results}}
From 456 positive {\sc secvtx} tags in 375 events in the lepton + jets sample corresponding to 695 pb$^{-1}$ we measure
\begin{equation}
\langle L_{2D}  \rangle = 0.581~\textrm{cm}
\end{equation}

\begin{figure}[ht]
  \begin{center}
    \includegraphics[width=0.8\textwidth,clip=]{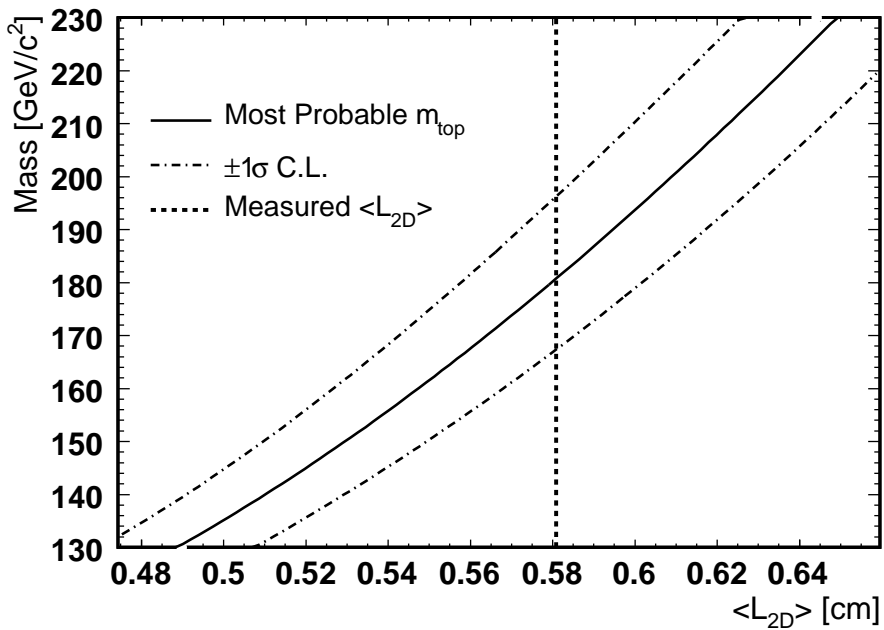}
    \caption{Most-probable (solid) and $\pm1\sigma$ (broken) $m_t$ curves as a function of mean transverse decay length.  Uncertainties are statistical only.  Measured mean transverse decay length is overlaid as dashed line.\label{result1}}
  \end{center}
\end{figure}

\noindent
We draw a vertical line at this $\langle L_{2D}  \rangle$ and read off the intersections with the most probable value and $\pm1\sigma$ confidence interval curves obtained above as illustrated in Fig.~\ref{result1} to extract a measurement of
\begin{equation}
 m_t =  180.7^{~+15.5}_{~-13.4}~(stat.)~\pm~8.6~(syst.)~\textrm{GeV}/c^2.  
\end{equation}

\begin{figure}[ht]
  \begin{center}
    \includegraphics[width=0.8\textwidth,clip=]{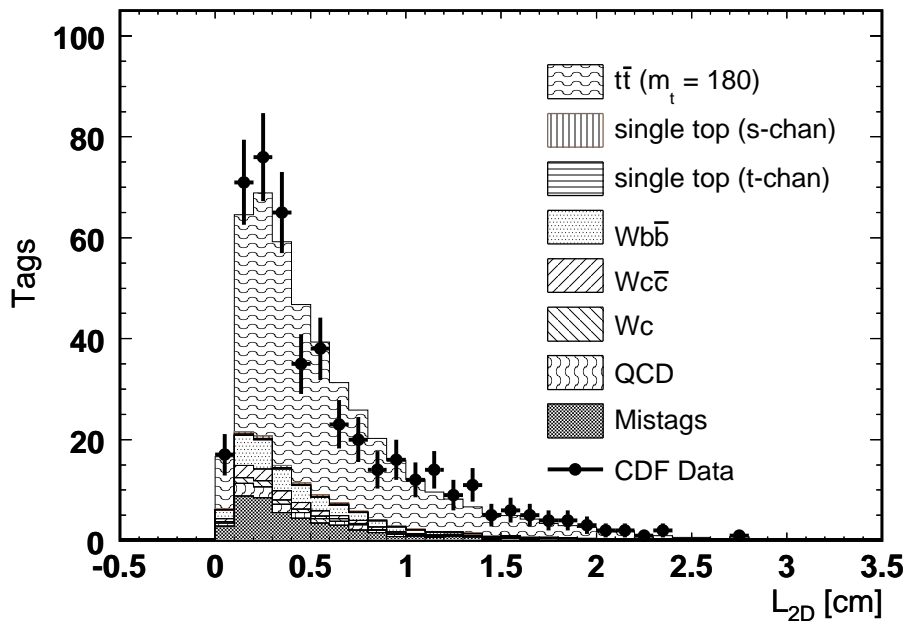}
    \caption{The $L_{2D}$ distribution of positive tags in selected events from which the mean transverse decay length used to measure top mass is extracted.  The points are the observed data.  Expected contributions from signal and background MC are displayed cumulatively in the histogram.  To facilitate the shape comparison, the MC is normalized to the observed data.\label{result2}}
  \end{center}
\end{figure}

\noindent
The $L_{2D}$ distribution of positive tags in selected events, from which the mean transverse decay length used to measure top mass is extracted, is shown together with expected contributions from signal and background MC overlaid in Fig.~\ref{result2}.   Reasonable agreement between data and MC is observed; a KS test yields a p-value of 16.7\%.

%
%


\section{Conclusion}
We have performed the first measurement of the top-quark mass using the decay length technique.  Using  695~pb$^{-1}$ data from Run II of the Tevatron we measure $m_t = 180.7^{~+15.5}_{~-13.4}~(stat.)~\pm~8.6~(syst.)$ GeV/$c^2$, consistent with the SM expectation as well as the combined result from of all existing measurements, $m_t = 171.4~\pm~2.1$ GeV/$c^2$~\cite{Brubaker:2006xn}.  Since it has negligible dependence on the jet energy scale and analyzes lepton plus three jet events which are not used by any other top mass measurement, the decay length technique is largely uncorrelated~\cite{correlation} with other methods.  Consequently, while the result presented in this paper is not a competitive measurement of $m_t$ by itself, it will help to reduce the overall uncertainty on the top mass when combined with other results.  In the combination referenced above, this measurement contributes at the few percent level to the overall combined result.

	This measurement is statistically limited and its dominant systematic uncertainties are likely reducible.  The precision of this measurement, therefore, will continue to improve over the course of Run II of the Tevatron.  The asymptotic performance of this technique at the Tevatron (and the LHC) are studied in detail in~\cite{Hill:2005zy}.  The current measurement establishes the technique and supports that paper's claim that the decay length method is a useful complement to existing measurements at the Tevatron and beyond. 	
	
\begin{acknowledgments}
We thank the Fermilab staff and the technical staffs of the participating institutions for their vital contributions. This work was supported by the U.S. Department of Energy and National Science Foundation; the Italian Istituto Nazionale di Fisica Nucleare; the Ministry of Education, Culture, Sports, Science and Technology of Japan; the Natural Sciences and Engineering Research Council of Canada; the National Science Council of the Republic of China; the Swiss National Science Foundation; the A.P. Sloan Foundation; the Bundesministerium f\"ur Bildung und Forschung, Germany; the Korean Science and Engineering Foundation and the Korean Research Foundation; the Particle Physics and Astronomy Research Council and the Royal Society, UK; the Institut National de Physique Nucleaire et Physique des Particules/CNRS; the Russian Foundation for Basic Research; the Comisi\'on Interministerial de Ciencia y Tecnolog\'{\i}a, Spain; the European Community's Human Potential Programme under contract HPRN-CT-2002-00292; and the Academy of Finland.
\end{acknowledgments}

\bibliography{lxy_cdf_prd}

\begin{thebibliography}{32}
\bibitem[{top()}]{toplife}
\bibinfo{note}{The lifetime of a top quark is estimated to be
  $\mathcal{O}(10^{-24})$ seconds.}
\bibitem[{coo()}]{coord}
\bibinfo{note}{We use a coordinate system where $\theta$\ is the polar angle to
  the proton beam, $\phi$\ is the azimuthal angle about this beam axis, and
  $\eta$ is the pseudorapidity defined as $-\ln\tan(\theta/2)$. The transverse
  momentum of a charged particle is denoted as $p_T \equiv p \sin\theta$. The
  analogously defined energy $E_T\equiv E \sin \theta$, is called transverse
  energy. Missing transverse energy, \MEt, is defined as the magnitude of
  $-\sum_i E_T^i \hat{n}_i$, where $\hat{n}_i$\ is a unit vector in the
  azimuthal plane that points from the beamline to the $i$th calorimeter
  tower.}
\bibitem[{jet()}]{jetcorr}
\bibinfo{note}{The $E_T$ values of jets are corrected for the effects of jet
  fragmentation, calorimeter non-uniformities and calorimeter absolute energy
  scale.}
\bibitem[{pur()}]{purity}
\bibinfo{note}{We ascertain the purity for the muon case by an independent
  measurement which uses the transverse momentum of the muon relative to the
  jet axis in a kinematic fit to distinguish between $b$ and non-$b$ jet
  contributions to the sample. The purity for the electron case is inferred
  since the data/MC ratio observed agrees with that obtained in the muon case.}
\bibitem[{sin()}]{singletop}
\bibinfo{note}{This difference comes from the t-channel production mechanism
  which is responsible for $\sim2/3$ of the single-top yield.}
\end{thebibliography}

\end{document}